\documentclass[letterpaper,twoside,10pt]{ieeeconf}
\usepackage{generic}
\usepackage{cite}
\usepackage{amsmath,amssymb,amsfonts}
\usepackage{mathrsfs}
\usepackage{algorithmic}
\usepackage{graphicx}
\usepackage{tikz}
\usepackage{pgfplots}
\pgfplotsset{compat=1.18}
\usepackage{empheq}
\usepackage{comment}
\newtheorem{theorem}{Theorem}
\newtheorem{proposition}{Proposition}
\newtheorem{corollary}{Corollary}
\newtheorem{property}{Property}
\newtheorem{remark}{Remark}
\newtheorem{example}{Example}
\newtheorem{definition}{Definition}
\newtheorem{ass}{Assumption}
\newtheorem{lem}{Lemma}

\newenvironment{Remark}{\begin{remark}}{\hfill $\bullet$ \end{remark}}
\newenvironment{Theorem}{\begin{theorem}}{\hfill $\square$ \end{theorem}}

\newenvironment{lemma}{\begin{lem}}{\hfill $\square$ \end{lem}}
\usepackage{algorithm,algorithmic}
\usepackage{hyperref}
\hypersetup{hidelinks=true}
\usepackage{textcomp}
\begin{document}
\title{\bf Spectral Boundary Observer for Counter-Flow Heat Exchangers}
\author{M. C. Belhadjoudja, M. Maghenem, E. Witrant
\thanks{M. C. Belhadjoudja is with the Department of Applied Mathematics, University of Waterloo, 200 University Avenue West, Waterloo, ON, Canada, N2L 3G1 (e-mail: m2camilb@uwaterloo.ca).}
\thanks{M. Maghenem and E. Witrant are with Universit\'e Grenoble Alpes, CNRS, Grenoble-INP, GIPSA-lab, F-38000, Grenoble, France (e-mail: mohamed.maghenem,emmanuel.witrant@gipsa-lab.fr).}}

\maketitle
\pagenumbering{gobble}

\begin{abstract}
We consider a system of two coupled first-order linear hyperbolic partial differential equations modeling heat transport in a counter-flow heat exchanger: one equation describes the transport of a hot fluid, and the other the transport of a cold fluid in the opposite direction. For this system, we design a boundary observer that uses only the temperature of the cold fluid measured at one boundary. Our approach is spectral: by assigning the spectrum of the operator governing the observation error dynamics to a prescribed region within the open left-half complex plane, we can freely tune the convergence rate of the observation error to zero in the $L^2$ norm. The main technical contribution is the proof that spectral stability, that is, the location of the spectrum in the open left-half plane, is equivalent to $L^2$ exponential stability of the origin for the observation error dynamics. This equivalence is established by showing that the operator governing the observation error dynamics satisfies the so-called spectral mapping property.
\end{abstract}

\section{Introduction}
Heat exchangers are devices designed to transfer thermal energy between two media, typically fluids, without direct mixing. They are ubiquitous across a broad spectrum of industrial sectors, including chemical processing, food and pharmaceutical production, oil and gas refining, and power generation~\cite{incropera,shah}. Their widespread deployment and the inherently slow dynamics they exhibit make the problems of real-time monitoring of distributed temperature profiles and their regulation toward desired references essential for ensuring performance, safety, and energy efficiency.

Counter-flow heat exchangers can be modeled by a system of two coupled first-order linear hyperbolic partial differential equations (PDE)s: one modeling the transport of the hot fluid, and the other one modeling the transport of the cold fluid in the opposite direction~\cite{bonne}. In this article, we consider this standard model and address the problem of observer design relying only on a boundary output, corresponding to the temperature of the cold fluid measured at one end of the exchanger.

Several approaches to observer design for hyperbolic systems, and, in particular, for the counter-flow heat-exchanger model, have been pursued in the literature. Lyapunov-based techniques have been employed in~\cite{castillo,zobiri} to establish $L^2$ exponential stability of the observation error dynamics. These approaches do not in general guarantee improved performance over the \textit{direct model}, which consists of a copy of the plant, without any correction. Another line of work exploits the 
port-Hamiltonian framework~\cite{kazaku}. However, as with Lyapunov-based methods, it does not allow the ability to tune the convergence rate of the observation error. The backstepping method has also been successfully applied to observer design for general hyperbolic systems, yielding observers with improved convergence rate~\cite{coron_backstepping,coron_2,ghousein}. However, this approach requires solving the so-called kernel PDEs, the complexity of which grows with the number of coupled equations and can become a significant obstacle for networks of PDEs, such as interconnections of heat exchangers \cite{networks}.

In this paper, we introduce a spectral methodology for boundary observer design for linear hyperbolic PDEs that avoids the limitations of the aforementioned approaches. By assigning the spectrum of the operator governing the observation errors to be within a prescribed region of the complex plane, the proposed method allows one to freely specify the convergence rate of the observation error to zero. While spectral methods have a long history in the control of parabolic PDEs, owing to the favorable spectral properties of the Laplacian, their extension to hyperbolic systems is considerably more delicate. The recent work~\cite{dus} has made it possible to apply spectral techniques to  boundary stabilization of certain classes of linear hyperbolic equations. However, the results in~\cite{dus} are restricted to full-state feedback control, and cannot be directly applied to observer design. The present work addresses this gap. The main technical challenge lies in establishing the equivalence between spectral stability (that is, the location of the entire spectrum in the open left-half plane) and exponential stability in the $L^2$ sense. While such  equivalence is classical in finite-dimensional systems, it fails to hold for general infinite-dimensional systems; see~\cite{counter_stab} for a counterexample involving a hyperbolic PDE. Our approach consists in proving that the operator governing the observation error satisfies the so-called \textit{spectral mapping property}~\cite{SMP_p,engel}, which enables to infer $L^2$ exponential stability directly from spectral stability.

The remainder of the paper is organized as follows. In Section~\ref{prob}, we introduce the system, the general form of our observer, and state our objective. In Section~\ref{main}, we state our main results: we first establish the equivalence between spectral stability and $L^2$ exponential stability, then design the observer's correction gains, and analyze stability of the observation errors. Our results are illustrated by numerical simulations in Section~\ref{simu}. Finally, Section~\ref{sec_conc} concludes the paper and discusses research perspectives. Due to space constraints, only proof sketches are provided in this paper, the complete proofs are available in \cite{complete}.

\textit{Notations.} We define $\mathbb{C}_{<0} := \{\lambda \in \mathbb{C} : \text{Re}(\lambda)<0\}$ where $\text{Re}(\lambda)$ denotes the real part of the complex number $\lambda$. The imaginary part of $\lambda$ is denoted $\text{Im}(\lambda)$. For a matrix $A\in \mathbb{C}^{n\times m}$, we denote by $|A|_2 := \sup_{|v|=1}|Av|$ its spectral norm. For a vector $v\in \mathbb{C}^n$, we denote by $|v| := \sqrt{\bar{v}^{\top}v}$ its Euclidean norm, where $\bar{v}$ denotes the conjugate of $v$. We let $L^2([0,1];\mathbb{C}^n)$ denote the Hilbert space of (equivalence classes of) Lebesgue measurable functions $v:[0,1]\to \mathbb{C}^n$ with $|v|_{L^2} := \left(\int_{0}^{1}|v(x)|^2\,dx\right)^{1/2} < +\infty$. Similarly, $H^1([0,1];\mathbb{C}^n)$ denotes the Sobolev space of functions in $L^2([0,1];\mathbb{C}^n)$ possessing a weak derivative $v'$ in $L^2([0,1];\mathbb{C}^n)$, equipped with the norm $|v|_{H^1} := (|v|_{L^2}^2 + |v'|_{L^2}^2)^{1/2}$. For $k\in \mathbb{N}$, a Hilbert space $\mathbb{H}$, and an interval $I\subset \mathbb{R}$, we denote by $\mathcal{C}^k(I;\mathbb{H})$ the space of functions $v:I\to \mathbb{H}$ that are $k$-times continuously differentiable with all derivatives up to order $k$ belonging to $\mathbb{H}$.

Given a Hilbert space $\mathbb{H} \neq \{0\}$ with norm $|\cdot |_{\mathbb{H}}$, we denote by $\mathcal{L}(\mathbb{H})$ the space of bounded linear operators $\mathcal{B}:\mathbb{H}\to \mathbb{H}$, i.e., with operator norm 
\begin{align*}
|\mathcal{B}|_{\mathbb{H}} := \sup\{|\mathcal{B}v|_{\mathbb{H}} : v\in \mathbb{H} \ \text{and} \ |v|_{\mathbb{H}}=1\} <+\infty.  
\end{align*}

The partial derivative of a function $(x,t)\mapsto z(x,t)$
with respect to $t$ is denoted $\partial_t z$, and the partial derivative with respect to $x$ is denoted $\partial_x z$. The derivative of a function $t\mapsto z(t)$ is denoted either $\dot{z}$ or $\frac{dz}{dt}$. The derivative of a function $x\mapsto z(x)$ is denoted either $z'$ or $\frac{dz}{dx}$.  

\section{Problem Formulation}\label{prob}

\subsection{System Description}

We consider the model of a counter-flow heat exchanger of the form
\begin{equation}
\label{eq1}
\left\lbrace
\begin{aligned}
&\partial_{t}T^{h} = -u_1 \partial_{x}T^{h} - c_1 (T^h-T^c),
\\
&\partial_{t}T^{c} = u_2 \partial_{x}T^{c} + c_2(T^h-T^c),
\end{aligned}
\right.
\end{equation}
where $T := [T^h \ T^c]^{\top} : [0,1]\times [0,+\infty)\to \mathbb{R}^2$ is the state, $x\in [0,1]$ is the space variable, $t\geq 0$ is the time variable, $u_{1}, u_{2} >0$ are constant transport velocities, and $c_{1}, c_{2}>0$ are constant coupling coefficients. The function $T^h$ (resp., $T^c$) is the temperature of the hot fluid (resp., of the cold fluid), $u_1$ (resp., $u_2$) is the flow rate of the hot fluid (resp., of the cold fluid), and $c_1$ and $c_2$ are the heat transfer coefficients.

The system \eqref{eq1} is subject to the boundary conditions
\begin{align}
T^h(0,t) = g^h(t), \ \ T^c(1,t) = g^c(t) \quad \forall t\geq 0, \label{bc_minus}
\end{align}
where $g^h,g^c\in \mathcal{C}^1(\mathbb{R}_{\geq 0};\mathbb{R})$. 

Finally, \eqref{eq1}-\eqref{bc_minus} are completed by the initial conditions
\begin{align}
T^h(x,0) = T_o^h(x), \quad T^c(x,0) = T_o^c(x) \quad \forall x\in [0,1], \label{init}
\end{align}
where $T_o^h,T_o^c\in H^1([0,1];\mathbb{R})$ verify the $0^{th}$-order compatibility conditions
\begin{align}
T_o^h(0) = g^h(0), \ \ T_o^c(1) = g^c(0). \label{compat_2}
\end{align}

Given the regularity assumptions on the boundary and initial conditions, as well as the compatibility conditions in \eqref{compat_2}, it can be shown using semigroup theory \cite[Chapter 5, Theorem 5.2]{pazy} that the system \eqref{eq1}-\eqref{init} admits a unique forward-complete classical solution in the following sense. 

\begin{definition}
A forward-complete classical solution to \eqref{eq1}-\eqref{init} is any function $T := [T^h \ T^c]^{\top}$ belonging to
$$\mathcal{C}^1(\mathbb{R}_{> 0};L^2([0,1];\mathbb{R}^{2}))\cap \mathcal{C}(\mathbb{R}_{\geq 0};H^1([0,1];\mathbb{R}^{2})),$$
that verifies \eqref{eq1} for all $x\in (0,1)$ and $t>0$, \eqref{bc_minus} for all $t\geq 0$, and \eqref{init} for all $x\in [0,1]$.
\hfill $\bullet$
\end{definition}

\subsection{Objective}

Our objective is to design an observer to estimate $T$ asymptotically at a prescribed convergence rate, using the output
\begin{align}
y(t) := T^c(0,t) \quad \forall t\geq 0. \label{output}
\end{align}
For this purpose, we assume that $\{u_i,c_i\}_{i\in \{1,2\}}$ and $\{g^i\}_{i\in \{h,c\}}$ are known, whereas $T_o^h$ and $T_o^c$ are unknown. To this end, we consider the observer
\begin{equation}\label{observer}
\left\lbrace
\begin{aligned}
\partial_t\hat{T}^h &= -u_1 \partial_x\hat{T}^h -c_1(\hat{T}^h-\hat{T}^c) + \kappa^h(x)(y-\hat{y}), \\
\partial_t\hat{T}^c &= u_2 \partial_x\hat{T}^c + c_2(\hat{T}^h-\hat{T}^c) + \kappa^c(x)(y-\hat{y}),
\end{aligned}
\right.
\end{equation}
where $\hat{T}:= [\hat{T}^h \ \hat{T}^c]^{\top}$ is the observer's state and $\hat{y}(t) := \hat{T}^c(0,t)$. 

The following boundary conditions are imposed
\begin{align}
\hat{T}^h(0,t) = g^h(t), \ \ \hat{T}^c(1,t) = g^c(t)\quad \forall t\geq 0. \label{obs_bc_minus}
\end{align}
Finally, the initial conditions are given by
\begin{align}
\hat{T}^h(x,0) = \hat{T}_o^h(x), \quad \hat{T}^c(x,0) = \hat{T}_o^c(x) \quad \forall x\in [0,1], \label{obs_init}
\end{align}
where $\hat{T}_o^h,\hat{T}_o^c\in H^1([0,1];\mathbb{R})$ verify 
\begin{align}
\hat{T}_o^h(0) = g^h(0), \ \  \ \hat{T}_o^c(1) = g^c(0).
\end{align}
Here, the gains $\kappa^h,\kappa^c\in H^{1}([0,1];\mathbb{C})$ are observer gains to be designed. These gains are complex valued as induced by the forthcoming spectral design approach. Accordingly, the estimates $\hat{T}^h$ and $\hat{T}^c$ are also complex valued.

The observation errors are denoted
\begin{align}
\tilde{T}^h := T^h - \hat{T}^h, \quad \tilde{T}^c := T^c - \hat{T}^c.
\end{align}
The errors $\tilde{T}^h$ and $\tilde{T}^c$ verify
\begin{equation}\label{observer_e}
\left\lbrace
\begin{aligned}
\partial_t\tilde{T}^h &= -u_1 \partial_x\tilde{T}^h -c_1(\tilde{T}^h-\tilde{T}^c) - \kappa^h(x)\tilde{T}^c(0), \\
\partial_t\tilde{T}^c &= u_2 \partial_x\tilde{T}^c + c_2(\tilde{T}^h-\tilde{T}^c) - \kappa^c(x)\tilde{T}^c(0),
\end{aligned}
\right.
\end{equation}
subject to the boundary conditions
\begin{align}
\tilde{T}^h(0,t) = \tilde{T}^c(1,t) = 0 \quad \forall t\geq 0, \label{err_bc_minus}
\end{align}
and to the initial conditions
\begin{align}
\tilde{T}^h(x,0) = \tilde{T}_o^h(x), \quad \tilde{T}^c(x,0) = \tilde{T}_o^c(x) \quad \forall x\in [0,1], \label{err_init}
\end{align}
where $\tilde{T}_o^h := T_o^h - \hat{T}_o^h$ and $\tilde{T}_o^c := T_o^c - \hat{T}_o^c$.

Given \eqref{observer_e}-\eqref{err_init}, the problem of observer design reduces to that of asymptotic stabilization of the origin $\{(\tilde{T}^h,\tilde{T}^c)=0\}$ via the design of $\kappa^h$ and $\kappa^c$. Specifically, given any desired convergence rate $\lambda_o>0$, the objective is to design $\kappa^h$ and $\kappa^c$ such that \eqref{observer_e}-\eqref{err_init} admits a unique forward-complete classical solution verifying the following property.

\begin{property}\label{prop1}
There exists $M_{\lambda_o}\geq 1$ such that, for all $t\geq 0$,
\begin{align}
|\tilde{T}(\cdot,t)|_{L^2([0,1];\mathbb{C}^2)} \leq M_{\lambda_o}|\tilde{T}_o|_{L^2([0,1];\mathbb{R}^{2})}\exp^{-\lambda_o t}, \label{decay_L2}
\end{align}
where $\tilde{T} := [\tilde{T}^h \ \tilde{T}^c]^{\top}$ and $\tilde{T}_o := [\tilde{T}_o^h \ \tilde{T}_o^c]^{\top}$.
\hfill $\bullet$
\end{property}

Since $T^h$ and $T^c$ are real valued, then \eqref{decay_L2} implies that
\begin{align}
|T(\cdot,t) - &\text{Re}(\hat{T}(\cdot,t))|_{L^2([0,1];\mathbb{R}^{2})} \nonumber \\
&\leq M_{\lambda_o}|\tilde{T}_o|_{L^2([0,1];\mathbb{R}^{2})}\exp^{-\lambda_o t} \quad \forall t\geq 0.
\end{align}
In other words, the real part of the observer's state converges exponentially to the system's state, at the prescribed rate $\lambda_o$, and in the sense of the $L^2$ norm.

\section{Main Results}\label{main}

\subsection{ Spectral Stability vs $L^2$ Stability}

To design $\kappa := [\kappa^h \ \kappa^c]^{\top}$, we express \eqref{observer_e}-\eqref{err_init} in the abstract form
\begin{equation}\label{abstract}
\left\lbrace
\begin{aligned}
\dot{z} &= (\mathcal{A} - \kappa\mathcal{C})z, \\
z(t=0) &= \tilde{T}_o,
\end{aligned}
\right.
\end{equation}
where $z:\mathbb{R}_{\geq 0}\to H^1([0,1];\mathbb{C}^{2})$, $t\mapsto z(t) := \tilde{T}(\cdot,t)$. To simplify the notations, we let
\begin{align}
\mathcal{H} := L^2([0,1];\mathbb{C}^{2}), \quad \mathcal{H}^1 := H^1([0,1];\mathbb{C}^{2}),
\end{align}
and $|\cdot|_{\mathcal{H}}$, $|\cdot|_{\mathcal{H}^1}$ be the corresponding norms, and $\langle\cdot,\cdot\rangle_{\mathcal{H}}$ the usual $L^2$ inner product.

The operator $\mathcal{A}:D(\mathcal{A})\to\mathcal{H}$, with domain
\begin{align}
D(\mathcal{A}) = \left\{z = \begin{bmatrix} z^h \\ z^c \end{bmatrix} \in \mathcal{H}^1 : z^h(0) = z^c(1) = 0\right\}, \label{domain_A}
\end{align}
is given by
\begin{align}
\mathcal{A}z := Uz_x + Mz, \label{A_def}
\end{align}
where 
\begin{align}
U := \begin{bmatrix} -u_1 & 0 \\ 0 & u_2 \end{bmatrix}, \quad
M := \begin{bmatrix} -c_1 & c_1 \\ c_2 & -c_2 \end{bmatrix}. \label{U_M_def}
\end{align}
The output operator $\mathcal{C}:D(\mathcal{A})\to\mathbb{C}$ is given by
\begin{align}
\mathcal{C}z := z^c(0). \label{C_def}
\end{align}

First, we show that \eqref{abstract} admits a unique forward-complete classical solution. To this end, we first need to recall the concept of strongly continuous semigroup generated by an operator, which, essentially, plays the role of the state-transition matrix for linear finite-dimensional systems \cite{pazy}.

\begin{definition}\label{def_semigroup}
Let $\mathbb{H}$ be a Hilbert space. A family $\{\mathcal{T}^t\}_{t\geq 0}$ of operators in $\mathcal{L}(\mathbb{H})$ is a strongly continuous semigroup on $\mathbb{H}$ if it verifies the following properties.
\begin{itemize}
\item $\mathcal{T}^0 = \mathcal{I}$.
\item $\forall t,\tau\geq 0$, $\mathcal{T}^{t+\tau} = \mathcal{T}^t\mathcal{T}^\tau$.
\item $\forall z_o\in\mathcal{H}$, $\lim_{t\to 0^+}\mathcal{T}^tz_o = z_o$.
\end{itemize}

Now, let $\mathcal{B}:D(\mathcal{B})\subset\mathbb{H}\to\mathbb{H}$. We say that $\mathcal{B}$ generates (or is the generator of) a strongly continuous semigroup $\{\mathcal{T}^t\}_{t\geq 0}\subset\mathcal{L}(\mathbb{H})$ if the following two conditions hold.
\begin{align}
&D(\mathcal{B}) = \left\{z\in\mathbb{H} : \lim_{t\to 0^+}\frac{\mathcal{T}^tz-z}{t} \text{ exists in } \mathbb{H}\right\}, \\
&\mathcal{B}z = \lim_{t\to 0^+}\frac{\mathcal{T}^tz-z}{t} \quad \forall z\in D(\mathcal{B}).
\end{align}
If $\mathcal{B}$ generates a strongly continuous semigroup $\{\mathcal{T}^t\}_{t\geq 0}$, then we write $\exp^{t\mathcal{B}} := \mathcal{T}^t$.
\hfill $\square$
\end{definition}

According to Lemma \ref{sol_li} in the Appendix, the existence of a unique forward-complete classical solution to \eqref{abstract} reduces to proving the following result (see Appendix \ref{semi_p} for a sketch of proof of Lemma \ref{semi}).

\begin{lemma}\label{semi}
Given any $\kappa \in \mathcal{H}^1$, the operator $\mathcal{A}-\kappa \mathcal{C}$ generates a strongly-continuous semigroup on $\mathcal{H}$. 
\end{lemma}

Then, and this is the main technical contribution of our work which was not needed for the stabilization problem in \cite{dus}, we show that Property \ref{prop1} can be ensured by designing $\kappa$ such that the spectrum of the generator $\mathcal{A}-\kappa \mathcal{C}$, denoted $\sigma(\mathcal{A}-\kappa \mathcal{C})$, lies in a desired subregion of the complex left-half plane. Such a result is non-trivial, since, as recalled in the introduction, spectral stability does not imply Lyapunov stability in general for infinite-dimensional systems.

Below, we recall the definition of spectrum of an operator. 

\begin{definition}\label{def_s}
Let $\mathbb{H}$ be a Hilbert space, and $\mathcal{B}:D(\mathcal{B})\subset\mathbb{H}\to\mathbb{H}$. The spectrum of $\mathcal{B}$ is defined by
\begin{align}
\sigma(\mathcal{B}) := \mathbb{C}\setminus\rho(\mathcal{B}),
\end{align}
where $\rho(\mathcal{B})$ is the resolvent set of $\mathcal{B}$, given by
\begin{align}
\rho(\mathcal{B}) := \left\{\lambda\in\mathbb{C} :  (\lambda\mathcal{I}-\mathcal{B})^{-1}\in\mathcal{L}(\mathbb{H})\right\}.
\end{align}
Finally, we define the point spectrum of $\mathcal{B}$ as
\begin{align}
\sigma_p(\mathcal{B}) := \left\{\lambda\in\sigma(\mathcal{B}) : \lambda\mathcal{I}-\mathcal{B} \text{ is not injective}\right\}.
\end{align}
In particular, $\lambda\in\sigma_p(\mathcal{B})$ if and only if there exists a nonzero function $v_\lambda\in D(\mathcal{B})$ such that $\mathcal{B}v_\lambda = \lambda v_\lambda$. The scalar $\lambda$ is called an eigenvalue of $\mathcal{B}$, and $v_\lambda$ is a corresponding eigenfunction.
\hfill $\square$
\end{definition}

Now, we can state our first main result.

\begin{Theorem}\label{thm1}
Let $\lambda_o>0$ and suppose that $\sigma(\mathcal{A}-\kappa \mathcal{C})\neq \emptyset$, where $\kappa\in \mathcal{H}^1$ and $\mathcal{A}$ and $\mathcal{C}$ are defined in \eqref{domain_A}-\eqref{C_def}. Then, Property~\ref{prop1} holds if and only if 
\begin{align}
\sigma(\mathcal{A}-\kappa\mathcal{C}) \subset (-\infty,-\lambda_o] + i\mathbb{R}. \label{obj_obs}
\end{align}
\end{Theorem}

When $\kappa = 0$, Theorem \ref{thm1} is a classical result from the theory of hyperbolic evolution operators \cite{SMP_p}. To the best of our knowledge, the case $\kappa \neq 0$ remained open. A sketch of the proof in this case is provided in Appendix \ref{thm_p}. The main idea is to show that the operator $\mathcal{A} - \kappa \mathcal{C}$ satisfies the so-called Spectral Mapping Property (SMP) \cite{SMP_p,engel}
\begin{align}
\sigma\!\left(e^{t(\mathcal{A}-\kappa\mathcal{C})}\right)\setminus\{0\}
= \overline{e^{\,\sigma(\mathcal{A}-\kappa\mathcal{C})t}}\setminus\{0\}, \label{SMP}
\end{align}
where $\overline{e^{\,\sigma(\mathcal{A}-\kappa\mathcal{C})t}}$ denotes the closure of $e^{\sigma(\mathcal{A}-\kappa\mathcal{C})t}$. This property links the spectrum of the generator to that of the associated semigroup, allowing us to derive an upperbound on 
$t \mapsto |e^{t(\mathcal{A}-\kappa \mathcal{C})}|_{\mathcal{L}(\mathcal{H})}$ from $\sigma(\mathcal{A}-\kappa \mathcal{C})$, and hence on $t \mapsto |z(t)|_{\mathcal{H}}$.

\subsection{Design of the Gain $\kappa$ and Observer Analysis}\label{k_design}

Having established Theorem~\ref{thm1}, we now design $\kappa$ to ensure~\eqref{obj_obs}. To this end, we first recall the following result, which is a direct consequence of \cite[Proposition 12]{dus} and of the stability of $\{z=0\}$ when $\kappa=0$.

\begin{lemma}[\cite{dus}]\label{lem_spec}
The following properties hold. 
\begin{enumerate}
    \item $\sigma(\mathcal{A}) = \sigma_p(\mathcal{A})$.
    \item $\sigma_p(\mathcal{A}) \subset \mathbb{C}_{\leq 0}$.
    \item In any bounded region of the complex plane, $\mathcal{A}$ admits finitely-many eigenvalues.
\end{enumerate}
\end{lemma}

In other words, in any bounded region of the complex plane, the operator $\mathcal{A}$ has only finitely many eigenvalues. Consequently, to achieve \eqref{obj_obs}, it suffices to shift only finitely many elements of $\sigma(\mathcal{A})$. This motivates the introduction of the shifted operator  
\[
\mathcal{A}_s := \mathcal{A} + \lambda_o \mathcal{I},
\]
with $D(\mathcal{A}_s) = D(\mathcal{A})$, whose spectrum satisfies  
\[
\sigma(\mathcal{A}_s) = \sigma(\mathcal{A}) + \lambda_o.
\]
In particular, $\sigma(\mathcal{A}_s) = \sigma_p(\mathcal{A}_s)$ and 
\[
\mathrm{Card}\{\sigma_p(\mathcal{A}_s)\cap \mathbb{C}_{\geq 0}\} < +\infty.
\]
Hence, to achieve \eqref{obj_obs}, we design $\kappa$ such that  
\begin{align}
\sigma(\mathcal{A}_s - \kappa \mathcal{C}) \subset \mathbb{C}_{<0}. \label{to_establish}
\end{align}
This reformulation of \eqref{obj_obs} has the advantage that \eqref{to_establish} can be established by analyzing the boundedness of solutions to the system $\dot{\tilde{z}} = (\mathcal{A}_s - \kappa \mathcal{C})\tilde{z}$ in $L^2$. Note that, since $\mathcal{A}_s - \kappa \mathcal{C}$ is simply a shifted version of $\mathcal{A} - \kappa \mathcal{C}$, it also generates a strongly continuous semigroup and satisfies the same SMP~\eqref{SMP}. Therefore, the boundedness of $\tilde{z}$ implies \eqref{to_establish}, which in turn ensures that \eqref{obj_obs} holds.


As a first step to establish \eqref{to_establish}, we need to compute the eigenvalues and eigenfunctions of $\mathcal{A}_s$. This is done in the following lemma, whose proof is in \cite{complete}.
\begin{lemma}\label{lem_eig}
If $\lambda \in \sigma (\mathcal{A}_s)$, then $f(\lambda)=0$, where 
\begin{align}
f(\lambda) :=&~ \frac{u_1}{c_1}\left(\theta_1(\lambda)\exp^{\theta_1(\lambda)}-\theta_2(\lambda)\exp^{\theta_2(\lambda)}\right) \nonumber \\
&+\left(\frac{\lambda-\lambda_o+c_1}{c_1}\right) \left(\exp^{\theta_1(\lambda)}-\exp^{\theta_2(\lambda)}\right),
\end{align}
with
\begin{align}
\theta_1(\lambda) &:= \frac{(u_2-u_1)(\lambda-\lambda_o)+u_2c_1-u_1c_2}{u_1u_2}, \\
\theta_2(\lambda) &:= -\frac{(\lambda-\lambda_o)^2+(\lambda-\lambda_o)(c_1+c_2)}{u_1u_2}.
\end{align}
Conversely, if $f(\lambda)=0$, then $\lambda\in \sigma(\mathcal{A}_s)$ if  
\begin{align}
v_{\lambda} := \begin{bmatrix}
v_{\lambda}^h & v_{\lambda}^c
\end{bmatrix}^{\top} \in H^1([0,1];\mathbb{C}^2)
\label{eigenf_1}
\end{align}
is non-null, where 
\begin{align}
v_{\lambda}^h(x) :=&~ \exp^{\theta_1(\lambda)x}-\exp^{\theta_2(\lambda)x}, \\
v_{\lambda}^c(x) :=&~ \frac{u_1}{c_1}\left(\theta_1(\lambda)\exp^{\theta_1(\lambda)x}-\theta_2(\lambda)\exp^{\theta_2(\lambda)x}\right) \nonumber \\
&~+\frac{\lambda-\lambda_o+c_1}{c_1}\left(\exp^{\theta_1(\lambda)x}-\exp^{\theta_2(\lambda)x} \right).\label{eigenf_3}
\end{align}
In which case, $v_\lambda$ is an eigenfunction associated to $\lambda$.
\end{lemma}

To each unstable eigenvalue $\lambda \in \sigma(\mathcal{A}_s)\cap \mathbb{C}_{\geq 0}$, we associate the generalized eigenspace 
\begin{align}
\text{Eig}^G(\mathcal{A}_s,\lambda) := \text{Ker}\left(\left(\lambda \mathcal{I}-\mathcal{A}_s\right)^{k_{\lambda}}\right),
\end{align}
where $k_\lambda$ is the algebraic multiplicity of $\lambda$. Then, following \cite{dus}, we introduce the finite-dimensional \textit{unstable subspace} 
\begin{align}
\mathcal{V} := \bigoplus_{\lambda \in \sigma(\mathcal{A}_s)\cap \mathbb{C}_{\geq 0}}\text{Eig}^G(\mathcal{A}_s,\lambda),
\end{align}
endowed with the $L^2$ inner product of $\mathcal{H}$. Here, $\bigoplus$ denotes the usual direct sum of subspaces of a linear space. 

In the sequel, we let $q\in \mathbb{N}$ be the dimension of $\mathcal{V}$, and $\{w_{i}\}_{i=1}^{q}$ be an orthonormal basis. We recall that such a basis can always be constructed from another basis $\{v_i\}_{i=1}^{q}$ that is not orthogonal by applying the Gram-Schmidt algorithm \cite{gram}. 

To design $\kappa$, we consider the \textit{shifted} system 
\begin{align}
\dot{\tilde{z}} = (\mathcal{A}_s-\kappa \mathcal{C})\tilde{z}, \quad \tilde{z}(0)=\tilde{z}_o \in D(\mathcal{A}_s),\label{shift_system}
\end{align}
and project it onto $\mathcal{V}$. To this end, we define the projection operator 
\begin{align}
\mathcal{P}:\mathcal{H}\to \mathcal{V}, \quad \mathcal{P}\tilde{z} := \sum_{i=1}^{q}z_i w_i, 
\end{align}
where 
\begin{align}
z_i := \langle \tilde{z},w_i\rangle_{\mathcal{H}} \quad \forall i\in \{1,...,q\}.
\end{align}
Applying $\mathcal{P}$ to both sides of \eqref{shift_system}, we obtain 
\begin{align}
\mathcal{P}\dot{\tilde{z}} = \mathcal{P}\mathcal{A}_s\tilde{z} - \tilde{z}^c(0)\mathcal{P}\kappa.\label{proj_1}
\end{align}
The left-hand side of \eqref{proj_1} can be expressed explicitly as 
\begin{align}
    \mathcal{P}\dot{\tilde{z}} = \sum_{i=1}^{q}\frac{d}{dt}\langle \tilde{z},w_i\rangle_{\mathcal{H}} w_i = \sum_{i=1}^{q}\dot{z}_i w_i. \label{pdot}
\end{align}
For each $k\in \{1,...,q\}$, taking the inner product of $\mathcal{P}\dot{\tilde{z}}$ with $w_k$, we conclude, from \eqref{pdot} and by orthonormality of $\{w_i\}_{i=1}^{q}$, that 
\begin{align}
\langle \mathcal{P}\dot{\tilde{z}},w_k\rangle_{\mathcal{H}} = \left\langle  \sum_{i=1}^{q}\dot{z}_i w_i,w_k\right\rangle_{\mathcal{H}} &= \sum_{i=1}^{q}\langle w_i,w_k\rangle_{\mathcal{H}} \dot{z_i} \nonumber \\
&= \dot{z}_k.\label{proj_2}
\end{align}
Hence, for each $k\in \{1,...,q\}$, we have, by combining \eqref{proj_1} and \eqref{proj_2}, 
\begin{align}
\dot{z}_k = \left\langle \mathcal{P}\mathcal{A}_s\tilde{z},w_k\right\rangle_{\mathcal{H}} - \left\langle \mathcal{P}\kappa ,w_k\right\rangle_{\mathcal{H}} \tilde{z}^c(0). \label{eg_k}
\end{align}
Furthermore, given that $\mathcal{V}$ is an invariant subspace of $\mathcal{A}_s$, the projection operator $\mathcal{P}$ commutes with $\mathcal{A}_s$. Hence, 
\begin{align}
\mathcal{P}\mathcal{A}_s\tilde{z} &= \mathcal{A}_s \mathcal{P}\tilde{z}= \mathcal{A}_s \left(\sum_{i=1}^{q}z_iw_i\right).
\end{align}
Taking the inner product with $w_k$, we obtain 
\begin{align}
\langle \mathcal{P}\mathcal{A}_s\tilde{z},w_k\rangle_{\mathcal{H}} = \sum_{i=1}^q \langle \mathcal{A}_s w_i,w_k\rangle_{\mathcal{H}} z_i.
\end{align}
As a result, \eqref{eg_k} becomes 
\begin{align*}
\dot{z}_k =  \sum_{i=1}^q \langle \mathcal{A}_s w_i,w_k\rangle_{\mathcal{H}} z_i - \left\langle \mathcal{P}\kappa ,w_k\right\rangle_{\mathcal{H}} \tilde{z}^c(0).
\end{align*}
Next, note that we have 
\begin{align}
\left\langle \mathcal{P}\kappa ,w_k\right\rangle_{\mathcal{H}} &= \left\langle \sum_{i=1}^{q}\left\langle\kappa ,w_i\right\rangle_{\mathcal{H}} w_i,w_k\right\rangle_{\mathcal{H}} \nonumber \\
&=\sum_{i=1}^{q}\left\langle \left\langle\kappa ,w_i\right\rangle w_i,w_k\right\rangle_{\mathcal{H}}=\left\langle \kappa , w_k\right\rangle_{\mathcal{H}}.
\end{align}
Consequently, we obtain 
\begin{align}
\dot{z}_k =  \sum_{i=1}^q \langle \mathcal{A}_s w_i,w_k\rangle_{\mathcal{H}} z_i - \left\langle \kappa , w_k\right\rangle_{\mathcal{H}} \tilde{z}^c(0).\label{249}
\end{align}
We will rewrite \eqref{249} in matrix form. To this end, we let
\begin{align}
Z := \begin{bmatrix}
    z_1 & z_2 & \hdots & z_q
\end{bmatrix}^{\top}, 
\end{align}
whose evolution is governed by 
\begin{align}
\dot{Z} = \mathsf{A}Z - \mathsf{K} \tilde{z}^c(0). \label{103}
\end{align}
Here, $\mathsf{A}\in \mathbb{C}^{q\times q}$ is the matrix
\begin{align}
[\mathsf{A}]_{k,j} := \langle \mathcal{A}_sw_i,w_k\rangle_{\mathcal{H}}. \label{A_def_3}
\end{align}
Moreover, $\mathsf{K}\in \mathbb{C}^{q\times 1}$ is the column vector 
$$ [\mathsf{K}]_{k} := \left\langle \kappa , w_k\right\rangle_{\mathcal{H}}.$$
The next step is to express $\tilde{z}^c(0)$ in a form that will allow us to design $\mathsf{K}$. The challenge is that $\tilde{z}^c(0)$ does not depend only on $Z$. Indeed, $\tilde{z}$ can be decomposed as 
\begin{align}
\tilde{z}(x) &= [\mathcal{P}\tilde{z}](x) + [(\mathcal{I}-\mathcal{P})\tilde{z}](x) \nonumber \\
&= \sum_{i=1}^{q}z_i w_i(x) + [(\mathcal{I}-\mathcal{P})\tilde{z}](x).
\end{align}
Here, $[\mathcal{P}\tilde{z}](x)$ denotes the function $\mathcal{P}\tilde{z}$ evaluated at $x$. The same notation is used for $[(\mathcal{I}-\mathcal{P})\tilde{z}](x)$. By letting $w_{i}^c := \begin{bmatrix} 0 & 1 \end{bmatrix} w_i$, we have
\begin{align*}
\tilde{z}^c(0) = \sum_{i=1}^{q} z_i w_{i}^c(0) + \begin{bmatrix} 0 & 1 \end{bmatrix} [(\mathcal{I}-\mathcal{P})\tilde{z}](0).
\end{align*}
As a result, \eqref{103} can be rewritten as 
\begin{align}
\dot{Z} = (\mathsf{A}-\mathsf{K}\mathsf{C})Z + \mathsf{K}\mathsf{T}, \label{122}
\end{align}
where 
\begin{align}
\mathsf{T} &:= - \begin{bmatrix} 0 & 1 \end{bmatrix} [(\mathcal{I}-\mathcal{P})\tilde{z}](0), \\
\mathsf{C} &:= \begin{bmatrix} w_{1}^c(0) & \hdots & w_{q}^c(0)
\end{bmatrix}. \label{C_def_new}
\end{align}

The objective is to design $\mathsf{K}$ so that the matrix $\mathsf{A}-\mathsf{K}\mathsf{C}$ is Hurwitz. To this end, we need the following result, whose proof is in Appendix \ref{obs_p}. 
\begin{lemma}\label{obs_lem}
The pair $(\mathsf{A},\mathsf{C})$ is observable.
\end{lemma}

Since $(\mathsf{A},\mathsf{C})$ is observable, we can select $\mathsf{K} := \begin{bmatrix}
    K_1 & K_2 & \hdots & K_q
\end{bmatrix}^{\top}$ such that $\mathsf{A}-\mathsf{K}\mathsf{C}$ is Hurwitz. Then, we design $\kappa $ such that
\begin{align}
\left\langle \kappa , w_i\right\rangle_{\mathcal{H}} = K_i \quad \forall i\in \{1,...,q\}.
\end{align}
In particular, since $\{w_i\}_{i=1}^{q}$ is orthonormal, we can let
\begin{align}
\kappa := \sum_{i=1}^{q}K_i w_i. \label{kappa12}
\end{align}

We can now state the following result, whose proof is in Appendix \ref{app_thm2}.
\begin{theorem}\label{thm2}
Consider the system \eqref{observer_e}-\eqref{err_init}. Let $\lambda_o>0$, and let the gains $\kappa^h$ and $\kappa^c$ be designed according to \eqref{kappa12}. Furthermore, suppose that $\sigma(\mathcal{A}-\kappa \mathcal{C})\neq \emptyset$, where $\mathcal{A}$ and $\mathcal{C}$ are defined in \eqref{domain_A}-\eqref{C_def}, and $\kappa = [\kappa^h \  \ \kappa^c]^{\top}$. Then, there exists a unique forward-complete classical solution $\tilde{T}$ to \eqref{observer_e}-\eqref{err_init} that verifies Property \ref{prop1}.
\hfill $\square$
\end{theorem}

\begin{Remark}
    In the stabilization problem addressed in \cite{dus}, the boundary control input depends only on the finite-dimensional projected state, so that a term analogous to $\mathsf{K}\mathsf{T}$ does not appear in the projected dynamics. Exponential stability of the finite-dimensional part then follows directly, and, since the remaining infinitely-many stable modes are left unchanged by the feedback, Property~\ref{prop1} holds without invoking Theorem~\ref{thm1} for $\kappa\neq 0$.
\end{Remark}

\section{Simulation Results}\label{simu}
In this section, we evaluate our results through numerical simulations performed in MATLAB.

We simulate the system \eqref{observer_e}–\eqref{err_init} both with $\kappa=0$ (direct model) and with $\kappa$ designed as in Theorem \ref{thm2}, for a prescribed convergence rate $\lambda_o$. The spatial domain $[0,1]$ is uniformly discretized into $N=200$ points. The transport terms are approximated using the standard first-order upwind finite difference method: a backward scheme for $\partial_x \tilde{T}^h$ and a forward scheme for $\partial_x \tilde{T}^c$. Time integration is carried out using an implicit Euler scheme with time step $\Delta t = 2.5\times 10^{-3}$. The initial observation errors are set to $\tilde{T}_o^h(x) = 8\sin(\pi x)$ and $\tilde{T}_o^c(x) = 6\sin(\pi(1-x))$. We take $u_1 = u_2 = c_1 = c_2 = 1$, and $\lambda_o \in \{3,5\}$.

Following \cite{dus}, we approximate the eigenvalues and corresponding eigenfunctions of $\mathcal{A}s = \mathcal{A} + \lambda_o \mathcal{I}$ with non-negative real parts by computing the eigenvalues and eigenvectors of the spatially discretized system. In particular, for $\lambda_o = 3$, we find $q=1$ unstable eigenvalue, whereas for $\lambda_o=5$, we find $q=9$ unstable eigenvalues. Once an orthonormal basis of the unstable subspace is computed, the projected matrices $\mathsf{A} \in \mathbb{C}^{q\times q}$ and $\mathsf{C} \in \mathbb{C}^{1\times q}$ are constructed according to \eqref{A_def_3} and \eqref{C_def_new}. The observability of $(\mathsf{A},\mathsf{C})$ (cf. Lemma~\ref{obs_lem}) is verified numerically. The gain $\mathsf{K} \in \mathbb{C}^{q\times 1}$, which ensures that $\mathsf{A}-\mathsf{K}\mathsf{C}$ is Hurwitz, is designed using LQR, with the state-cost weighted matrix $Q=(\lambda_o+2)^2I_{q}$ where $I_q$ is the identity matrix of dimension $q$, and the input-cost weighted matrix $R=1$. The observer gain is constructed as in \eqref{kappa12}.

\begin{figure}
    \centering
    \includegraphics[width=\linewidth]{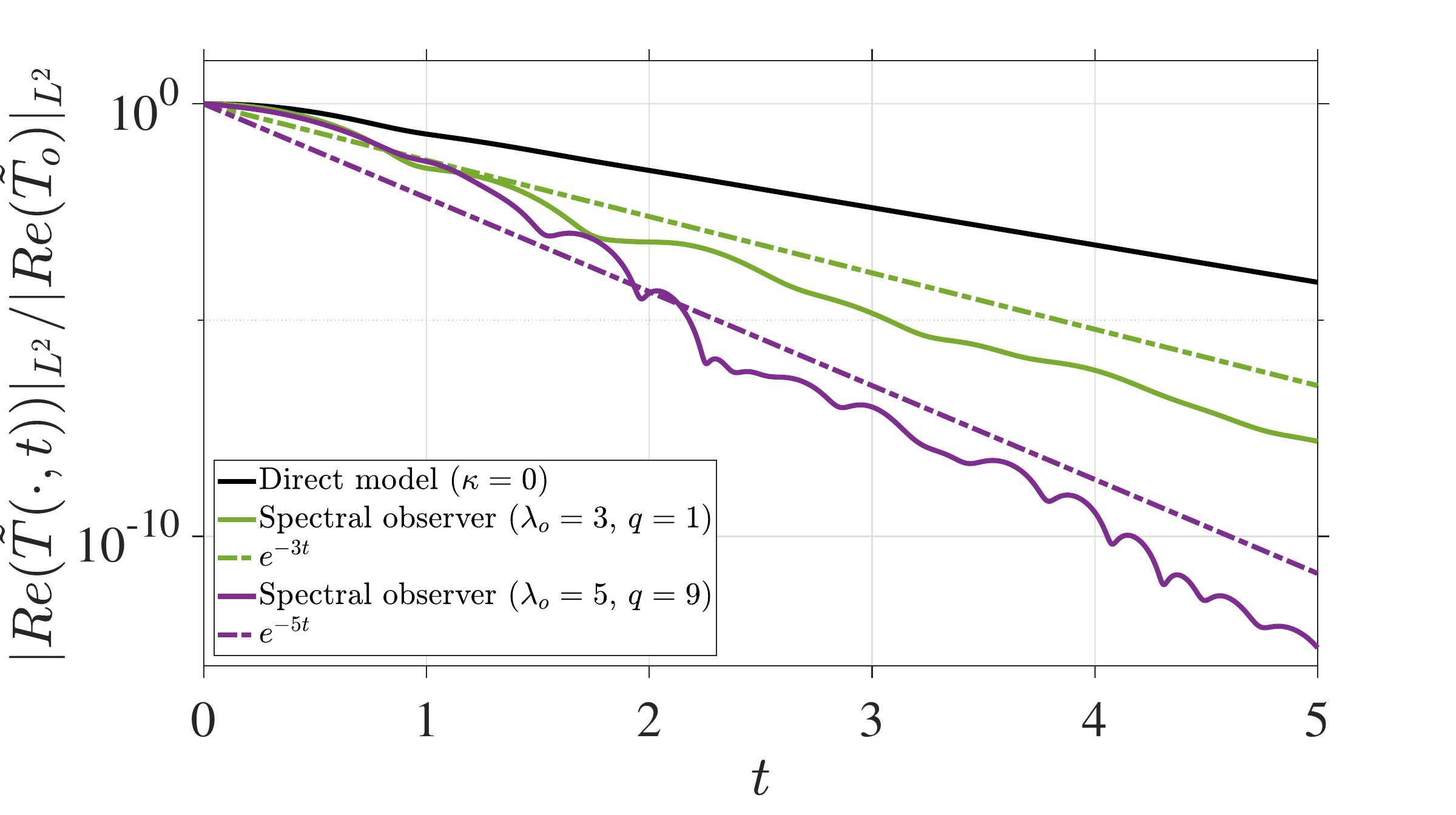}
    \caption{The scaled $L^2$ norm of the observation error for $\kappa=0$ vs. for $\kappa$ designed as in \eqref{kappa12} with $\lambda_o\in \{3,5\}$.}
    \label{fig1}
\end{figure}

\begin{figure}
    \centering
    \includegraphics[width=\linewidth]{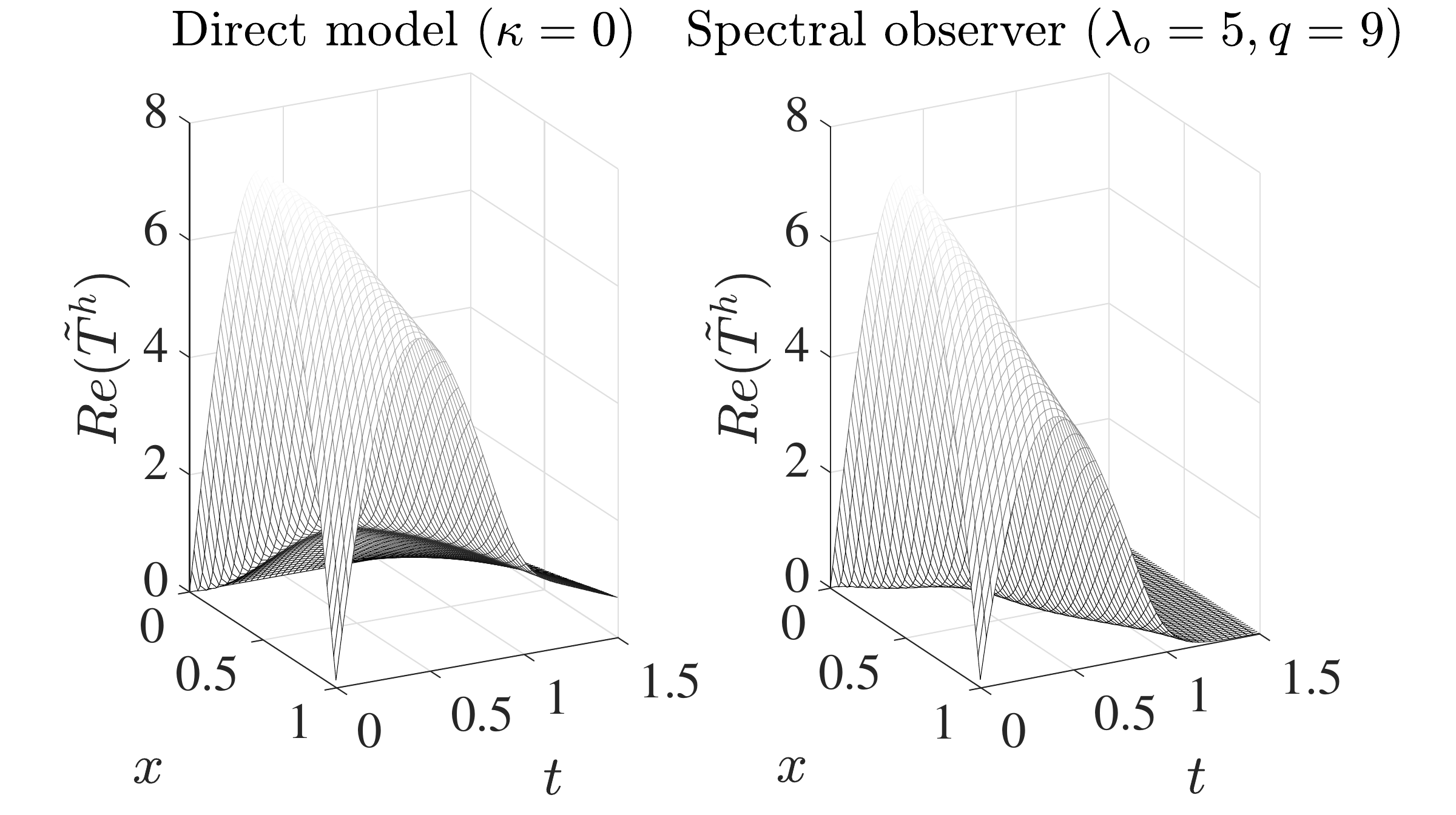}
    \caption{$\text{Re}(\tilde{T}^h)$ with the direct model (left) vs. with the spectral observer (right).}
    \label{fig2}
\end{figure}

\begin{figure}
    \centering
    \includegraphics[width=\linewidth]{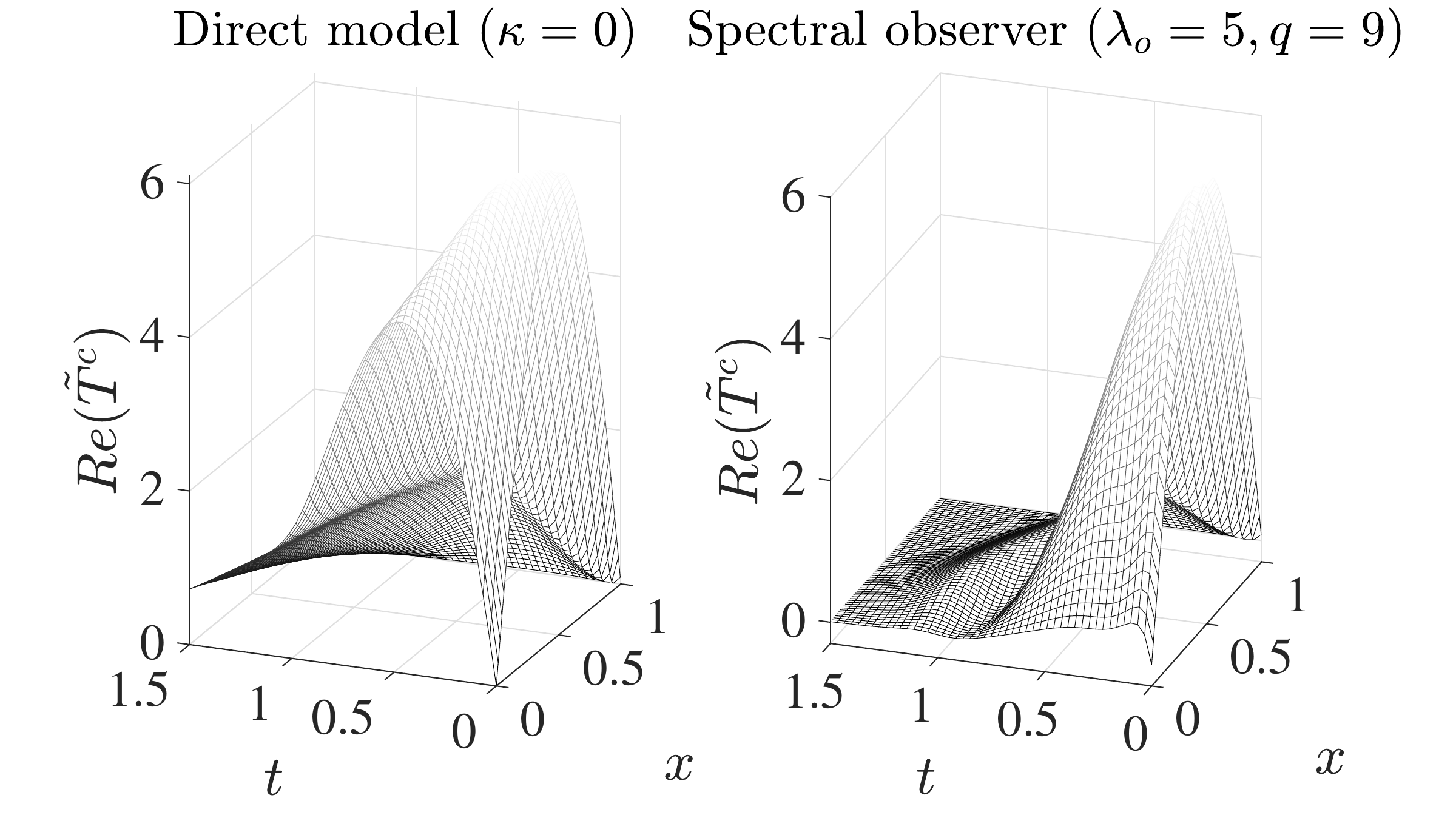}
    \caption{$\text{Re}(\tilde{T}^c)$ with the direct model (left) vs. with the spectral observer (right).}
    \label{fig3}
\end{figure}

In Figure \ref{fig1}, we plot $t\mapsto |\text{Re}(\tilde{T}(\cdot,t))|{L^2}/|\text{Re}(\tilde{T}_o)|{L^2}$, comparing the direct model ($\kappa = 0$) and the spectral observer for $\lambda_o\in \{3,5\}$. As expected, the convergence rate of the observation error to zero is higher with the spectral observer than with the direct model. Moreover, the convergence rate for $\lambda_o=5$ is higher than that for $\lambda_o=3$, confirming that the convergence rate can be improved by appropriately choosing $\lambda_o$. The dashed lines represent the functions $t\mapsto \exp(-3t)$ and $t\mapsto \exp(-5t)$. As expected, an overshoot can be observed, which corresponds to the constant $M_{\lambda_o}$ in Property \ref{prop1}. The superior performance of the spectral observer over the direct model can also be seen in Figures \ref{fig2} and \ref{fig3}, where we plot $\text{Re}(\tilde{T}^h)$ and $\text{Re}(\tilde{T}^c)$, respectively, for both the direct model and the spectral observer. Note that the orientations of Figures \ref{fig2} and \ref{fig3} are intentionally different to better illustrate the performance of the spectral observer compared to the direct model.

\section{Conclusion and Research Perspectives}\label{sec_conc}

In this paper, we have proposed a spectral approach to boundary observer design for a counter-flow heat exchanger model. The convergence rate of the $L^2$ norm of the observation error to zero can be freely prescribed by assigning the spectrum of the operator governing the observation error dynamics into a given subregion of the complex plane. A first perspective concerns the combination of the proposed observer with a boundary feedback controller that would actuate the flow rates $u_1$ and $u_2$ in order to regulate the temperature profile. When the flow rates $u_1$ and $u_2$ serve as control inputs, they become time-varying. The model \eqref{eq1} then involves time-dependent transport velocities, and the operator $\mathcal{A}$ becomes time-dependent as well. In this setting, the spectral analysis underpinning our approach, in particular, the spectral mapping property and the decomposition into stable and unstable subspaces, no longer applies directly, and must be adapted. Another perspective would be to attenuate the overshoot observed in the simulations. For instance, the overshoot attenuation problem is well studied in the high-gain observer literature \cite{khalil}. A further direction would be to study an adaptive redesign of our observer to handle cases where $c_1$ and $c_2$ are unknown. Although classical results on adaptive observer design for infinite-dimensional systems are available \cite{curtain}, their adaptation to our setting does not seem straightforward. Finally, it would be of interest to study the robustness of the observer to uncertainties in the eigenvalues. Indeed, the roots of the characteristic equation cannot be obtained in closed form in general and must be approximated numerically

\appendix

\subsection{Useful results}\label{defs}
We first recall the definition of strongly continuous semigroup generated by an operator.

Strongly continuous semigroups allow us to solve abstract Cauchy problems such as \eqref{abstract}, as shown in the following lemma; see \cite[Chapter 1, Theorem 2.4]{pazy}.

\begin{lemma}\label{sol_li}
Let $\mathbb{H}$ be a Hilbert space, and $\mathcal{B}:D(\mathcal{B})\subset\mathbb{H}\to\mathbb{H}$ be the generator of a strongly continuous semigroup $\{\exp^{t\mathcal{B}}\}_{t\geq 0}\subset\mathcal{L}(\mathbb{H})$. If $z_o\in D(\mathcal{B})$, then there exists a unique classical solution to the Cauchy problem
\begin{equation}
\left\lbrace
\begin{aligned}
\dot{z} &= \mathcal{B}z, \\
z(t=0) &= z_o.
\end{aligned}
\right.
\end{equation}
This solution is given by
\begin{align}
z(t) = \exp^{t\mathcal{B}}z_o \quad \forall t\geq 0. \label{semi_form}
\end{align}
\end{lemma}

The following result allows us to verify that an operator generates a strongly continuous semigroup. 

\begin{lemma}[Lumer-Phillips \cite{pazy}]\label{lumer}
Let $\mathbb{H}$ be a Hilbert space. An operator $\mathcal{B}:D(\mathcal{B})\subset\mathbb{H}\to\mathbb{H}$ generates a strongly continuous semigroup on $\mathbb{H}$ if and only if the following conditions hold.
\begin{enumerate}
\item $D(\mathcal{B})$ is dense in $\mathbb{H}$.
\item $\mathcal{B}$ is closed.
\item Dissipativity: There exists $\zeta \in \mathbb{R}$ such that
\begin{align}
\text{Re}\langle z, \mathcal{B}z\rangle_{\mathbb{H}} \leq \zeta |z|_{\mathbb{H}}^2 \quad \forall z \in D(\mathcal{B}).
\end{align}
\item The resolvent set of $\mathcal{B}$ is not empty.
\end{enumerate}
\end{lemma}

For strongly continuous semigroups, a characterization of the resolvent set (and thus, of the spectrum) is given by the following lemma. 

\begin{lemma}[\cite{gp_p1,gp_p2}]\label{GP}
Let $\mathbb{H}$ be a Hilbert space, and $\mathcal{B}:D(\mathcal{B})\subset\mathbb{H}\to\mathbb{H}$ be the generator of a strongly continuous semigroup $\{\exp^{t\mathcal{B}}\}_{t\geq 0}\subset\mathcal{L}(\mathbb{H})$. Given $\lambda\in\mathbb{C}$ and $t>0$, we have $\exp^{\lambda t}\in\rho\left(\exp^{t\mathcal{B}}\right)$ if and only if
\begin{align}
&\{\lambda + i2\pi t^{-1}j : j\in\mathbb{Z}\} \subset \rho(\mathcal{B}), \label{cond_1}\\
&\sup_{j\in\mathbb{Z}}\left|\mathcal{R}\left(\lambda + i2\pi t^{-1}j,\mathcal{B}\right)\right|_{\mathcal{L}(\mathbb{H})} < \infty, \label{cond_2}
\end{align}
where, for any $\mu\in\rho(\mathcal{B})$, $\mathcal{R}(\mu,\mathcal{B}) := (\mu\mathcal{I}-\mathcal{B})^{-1}$.
\end{lemma}

\subsection{Sketch of proof of Lemma \ref{semi}}\label{semi_p}

We verify the four conditions of the Lumer-Phillips theorem (cf. Lemma~\ref{lumer}) for $\mathcal{B}:=\mathcal{A}-\kappa\mathcal{C}$.

The \textit{density} of $D(\mathcal{A}-\kappa \mathcal{C})=D(\mathcal{A})$ in $\mathcal{H}$ is in~\cite{dus}.

For \textit{closedness}, let $(z_k)_{k\in\mathbb{N}}\subset D(\mathcal{A})$ with $z_k\to z$ and $(\mathcal{A}-\kappa\mathcal{C})z_k\to y$ in $\mathcal{H}$. The key step is to show that $(z_k^c(0))_{k\in\mathbb{N}}$ is Cauchy in $\mathbb{C}$. Letting $w_k := (\mathcal{A}-\kappa\mathcal{C})z_k$ and integrating from $x$ to $1$, we obtain $z_k^c(x) = \alpha_k^c(x) - z_k^c(0)\tilde{K}^c(x)$, where $\tilde{K}^c(x):=\int_x^1 u_2^{-1}\kappa^c(\xi)d\xi$ and $(\alpha_k^c)_{k\in \mathbb{N}}$ is a Cauchy sequence in $L^2$ that depends on $(w_k)_{k\in \mathbb{N}}$ and $(z_k)_{k\in \mathbb{N}}$. Since $(z_k)_{k\in \mathbb{N}}$ and $(w_k)_{k\in \mathbb{N}}$ are convergent (hence Cauchy) in $\mathcal{H}$, the sequence $(\alpha_k^c)_{k\in \mathbb{N}}$ is Cauchy in $L^2$. If $\kappa^c\not\equiv 0$, then $|\tilde{K}^c|_{L^2}>0$, so for all $k,\ell\in\mathbb{N}$,
\begin{align*}
|z_k^c(0)-z_\ell^c(0)|\leq \frac{|\alpha_k^c-\alpha_\ell^c|_{L^2}+|z_k^c-z_\ell^c|_{L^2}}{|\tilde{K}^c|_{L^2}},
\end{align*}
proving $(z_k^c(0))_{k\in \mathbb{N}}$ is Cauchy. If $\kappa^c\equiv 0$ but $\kappa^h\not\equiv 0$, the same argument applies to the $z^h$ component. The convergence of $(z_k^c(0))_{k\in \mathbb{N}}$ then implies convergence of $((d/dx)z_k)$ in $\mathcal{H}$, hence $z_k\to z$ in $\mathcal{H}^1$, and we conclude that $z\in D(\mathcal{A})$ and $(\mathcal{A}-\kappa\mathcal{C})z=y$.

For \textit{dissipativity}, integration by parts gives
\begin{align}
\hspace{-0.2cm}\text{Re}\langle z, \mathcal{A}z\rangle\leq&~ -\frac{u_1}{2}|z^h(1)|^2-\frac{u_2}{2}|z^c(0)|^2+|M|_2|z|_{\mathcal{H}}^2, \label{A_bound}
\end{align}
while the Cauchy-Schwarz and Young's inequalities yield
\begin{align}
|\text{Re}\langle z,\kappa\mathcal{C}z\rangle_{\mathcal{H}}|\leq \delta|z^c(0)|^2+\frac{|\kappa|_{\mathcal{H}}^2}{4\delta}|z|_{\mathcal{H}}^2 \ \ \forall \delta>0. \label{kC_bound}
\end{align}
Choosing $\delta = u_2/2$ in \eqref{kC_bound} and combining with \eqref{A_bound} gives
\begin{align}
\text{Re}\langle z, (\mathcal{A}-\kappa\mathcal{C})z\rangle_{\mathcal{H}} \leq \left(|M|_2+\frac{|\kappa|_{\mathcal{H}}^2}{u_2}\right)|z|_{\mathcal{H}}^2. \label{diss_final}
\end{align}

For the \textit{non-emptiness of} $\rho(\mathcal{A}-\kappa\mathcal{C})$, we reformulate the resolvent equation $(\lambda\mathcal{I}-(\mathcal{A}-\kappa\mathcal{C}))z=f$ for some $\lambda \in \mathbb{R}_{>0}$ as a fixed-point problem on $\mathcal{X}:=L^2([0,1];\mathbb{C})\times L^2([0,1];\mathbb{C})\times \mathbb{C}$, of the form $\mathcal{T}(z^h,z^c,z^c(0))=(z^h,z^c,z^c(0))$ for some operator $\mathcal{T} : \mathcal{X}\to \mathcal{X}$. Then, the operator $\mathcal{T}=\mathscr{L}+\mathscr{F}$ decomposes into a homogeneous part $\mathscr{L}$ (that is independent of $f$) and an $f$-dependent part $\mathscr{F}$. We show that $|\mathscr{L}|_{\mathcal{L}(\mathcal{X})}\leq C_M/\sqrt{\lambda}$ for a constant $C_M>0$ independent of $\lambda$. Taking $\lambda > C_M^2$, \cite[Theorem~10.13]{rudin} gives $|\mathscr{L}|_{\mathcal{L}(\mathcal{X})}<1$, so $\mathcal{I}-\mathscr{L}$ is invertible and the resolvent is bounded.

\subsection{Sketch of Proof of Theorem \ref{thm1}}\label{thm_p}
The key tool to address the case $\kappa \neq 0$ is Lemma \ref{GP} in Appendix \ref{defs}. Indeed, the inclusion $\sigma\left(\exp^{t(\mathcal{A}-\kappa\mathcal{C})}\right)\setminus\{0\} \subset  \overline{\exp^{\sigma(\mathcal{A}-\kappa\mathcal{C})t}}\setminus\{0\}$ follows from the spectral inclusion theorem~\cite[Chapter~IV]{engel}. For the reverse inclusion, we use Lemma~\ref{GP} and proceed by contraposition. Given $\lambda\in \mathbb{C}$ and $t>0$ such that $\exp^{\lambda t}\notin \overline{\exp^{\sigma(\mathcal{A}-\kappa\mathcal{C})t}}$, condition~\eqref{cond_1} follows from the $2\pi i$-periodicity of the exponential function. For condition~\eqref{cond_2}, we define, for each $j\in \mathbb{Z}$, the function
\begin{align}
\varphi_j := (\lambda+i2\pi t^{-1}j-(\mathcal{A}-\kappa\mathcal{C}))^{-1}f, \label{reso}
\end{align}
where $f\in \mathcal{H}$. The objective is to show that $|\varphi_j|_{\mathcal{H}}$ is bounded by a constant that is independent of $j$. To this end, we first note that \eqref{reso} can be rewritten as a differential equation on $\varphi_j$. Separating real and imaginary parts, we obtain, for each $j\in \mathbb{Z}$, a differential equation for $\Phi_j := [\text{Re}(\varphi_j)\ \text{Im}(\varphi_j)]^{\top}$, with some coefficients multiplying the rotation matrix
$\mathcal{J}:=\left[\begin{smallmatrix}0&1\\-1&0\end{smallmatrix}\right]$. These coefficients are shown to grow unboundedly as $|j|\to\infty$. Hence, we perform a change of variables that transform the system into one with uniformly bounded coefficients. After solving explicitly the resulting system, and analyzing the solution, we are able to conclude on the uniform boundedness of $\varphi_j$, and \textit{a fortiori}, \eqref{SMP} holds. 

Now that \eqref{SMP} has been established, we can conclude the proof of Theorem \ref{thm1}. Specifically, by \cite[Proposition~2.2.15]{marius}, if \eqref{obj_obs} holds, then there exists $M_{\lambda_o}\geq 1$ such that
$$|\exp^{t(\mathcal{A}-\kappa\mathcal{C})}|_{\mathcal{L}(\mathcal{H})}\leq M_{\lambda_o}\exp^{-\lambda_o t} \quad \forall t\geq 0.$$
Hence, \eqref{decay_L2} follows from Lemma \ref{sol_li}. The necessity of \eqref{obj_obs} for Property \ref{prop1} to hold is also a direct consequence of \cite[Proposition~2.2.15]{marius}.

\subsection{Proof of Lemma \ref{obs_lem}}\label{obs_p}
According to the Hautus test \cite{sontag}, it is enough to show that $\text{Ker}(\lambda I-\mathsf{A}) \cap \text{Ker}(\mathsf{C}) = \{0\} \quad \forall \lambda \in \mathbb{C}$, which can be done by proving that 
\begin{align}
\text{Ker}(\lambda \mathcal{I}-\mathcal{A}_s) \cap \text{Ker}(\mathcal{C}) = \{0\} \quad \forall \lambda \in \mathbb{C}. \label{120}
\end{align}
To establish \eqref{120}, we first recall that, if $\lambda\in \sigma(\mathcal{A}_s)$ and $v_{\lambda}$ is a corresponding eigenfunction defined by \eqref{eigenf_1}-\eqref{eigenf_3}, then 
\begin{align}
\text{Ker}(\lambda \mathcal{I}-\mathcal{A}_s) = \text{span}(v_\lambda).
\end{align}
Otherwise, $\text{Ker}(\lambda \mathcal{I}-\mathcal{A}_s)$ would reduce to zero, and \eqref{120} would hold trivially. Hence, we need to prove that the only function $\gamma v_{\lambda}$, with $\gamma\in \mathbb{C}$, that belongs to $\text{Ker}(\mathcal{C})$, is the zero function corresponding to $\gamma=0$. To do so, we note that if $\gamma v_{\lambda}\in \text{Ker}(\mathcal{C})$, then $0=\gamma \mathcal{C}v_{\lambda} = \gamma v_{\lambda}^c(0)$. In particular, we necessarily have either $\theta_1(\lambda) = \theta_2(\lambda)$ or $\gamma=0$. In the former case, we would have $v_{\lambda} = 0$, which would contradict the fact that $v_{\lambda}$ is an eigenfunction. Hence, we must have $\gamma=0$, which concludes the proof.

\subsection{Proof of Theorem \ref{thm2}}\label{app_thm2}

The existence of a unique forward-complete classical solution to \eqref{observer_e}-\eqref{err_init} follows from Lemmas \ref{semi} and \ref{sol_li}. To show that Property \ref{prop1} holds, we analyze the system
\begin{equation*}
\left\lbrace 
\begin{aligned}
\dot{Z} &= (\mathsf{A}-\mathsf{K}\mathsf{C})Z+\mathsf{K}\mathsf{T}, \\
(\mathcal{I}-\mathcal{P})\dot{\tilde{z}} &= (\mathcal{I}-\mathcal{P})\mathcal{A}_s\tilde{z} - (\mathcal{I}-\mathcal{P})\kappa \tilde{z}^c(0).
\end{aligned}
\right.
\end{equation*}
To this end, we first note that, since $\kappa \in \mathcal{V}$, then $(\mathcal{I}-\mathcal{P})\kappa = 0$. Hence, we have 
\begin{align}
(\mathcal{I}-\mathcal{P})\dot{\tilde{z}} = (\mathcal{I}-\mathcal{P})\mathcal{A}_s \tilde{z}.
\end{align}
Next, given that $\mathcal{A}_s$ commutes with $\mathcal{P}$, we can write 
\begin{align}
(\mathcal{I}-\mathcal{P})\dot{\tilde{z}} = \mathcal{A}_s (\mathcal{I}-\mathcal{P})\tilde{z}.\label{126}
\end{align}
Let $\mathcal{A}_s\big|_{\mathcal{V}'}$ be the restriction of $\mathcal{A}_s$ to $\mathcal{V}'$, where $\mathcal{V} \oplus \mathcal{V}' = \mathcal{H}$. Furthermore, define $\xi(x,t) := \big[(\mathcal{I}-\mathcal{P})\tilde{z}](x,t)$ and denote $\xi(t) := \xi(\cdot,t)$. Then, \eqref{126} can be rewritten as 
\begin{align}
\dot{\xi} = \mathcal{A}_s \big|_{\mathcal{V}'}\xi.
\end{align}
It has been shown in \cite{dus} that $\mathcal{A}_s \big|_{\mathcal{V}'}$ verifies the same SMP property as in \eqref{SMP}. Hence, since $\sigma (\mathcal{A}_s\big|_{\mathcal{V}'}) \subset \mathbb{C}_{<0}$, we conclude on the existence of $C_1\geq 0$ and $\sigma_1>0$ such that 
\begin{align}
|\xi(t)|_{\mathcal{H}} \leq C_1 |\xi(0)|_{\mathcal{H}}\exp^{-\sigma_1 t} \quad \forall t\geq 0. \label{xi_decay}
\end{align}
Now, using \eqref{xi_decay}, we will prove that $Z$ is bounded. To do so, we start by integrating \eqref{122} to obtain 
\begin{align*}
Z(t) = \exp^{t(\mathsf{A}-\mathsf{K}\mathsf{C})}Z(0) + \int_{0}^{t}\exp^{(t-\tau)(\mathsf{A}-\mathsf{K}\mathsf{C})}\mathsf{K}\mathsf{T}(\tau)d\tau.
\end{align*}
Since $\mathsf{A}-\mathsf{K}\mathsf{C}$ is Hurwitz, we conclude on the existence of $C_2,\sigma_2>0$ such that 
\begin{align}
\left| \exp^{(t-\tau)(\mathsf{A}-\mathsf{K}\mathsf{C})}\right|_2 \leq C_2 \exp^{-\sigma_2(t-\tau)}.
\end{align}
Hence, we have, for all $t\geq 0$, 
\begin{align}
|Z(t)|
\leq&~ C_2 |Z(0)|\exp^{-\sigma_2t}\nonumber \\
&~+C_2\mathsf{K}\left| \mathsf{T}\right|_{L^2(0,+\infty)}\sqrt{\frac{1-\exp^{-2\sigma_2t}}{2\sigma_2}}.
\end{align}
As a result, boundedness of $Z$ reduces to
\begin{align}
\left|\mathsf{T}\right|_{L^2(0,+\infty)} < +\infty. \label{to_show_T}
\end{align}
We claim that \eqref{to_show_T} is a consequence of \eqref{xi_decay}. Indeed, note that $\xi$ verifies $\xi_t = \mathcal{A}_s\xi=\mathcal{A}\xi + \lambda_o\xi$. Hence, 
\begin{align}
\frac{d}{dt}|\xi|_{\mathcal{H}}^2 =&~ 2\text{Re}\langle \mathcal{A}\xi,\xi\rangle_{\mathcal{H}}+\lambda_o|\xi|_{\mathcal{H}}^2 \nonumber \\
\leq&~ \left(2|M|_2+\lambda_o\right)|\xi|_{\mathcal{H}}^2-u_{2}\mathsf{T}^2, \label{276}
\end{align}
As a result, integrating both sides of \eqref{276} with respect to $t$, we obtain 
\begin{align*}
|\mathsf{T}|_{L^2(0,+\infty)}^2\leq \frac{2|M|_2+\lambda_o}{u_{2}}\int_{0}^{+\infty}|\xi(t)|_{\mathcal{H}}^2dt+|\xi(0)|_{\mathcal{H}}^2. 
\end{align*}
Using \eqref{xi_decay}, we conclude on \eqref{to_show_T}.

Finally, by the boundedness of $|\xi|_{\mathcal{H}}$ and $\left|Z\right|$, we conclude on the boundedness of $|\tilde{z}|_{\mathcal{H}}$. Consequently, \eqref{to_establish} must hold, since otherwise, there would exist a solution to $\dot{\tilde{z}}=(\mathcal{A}_s-\kappa \mathcal{C})\tilde{z}$ whose spatial $L^2$ norm is unbounded. Finally, given that \eqref{to_establish} holds, \eqref{obj_obs} also holds, which implies that Property \ref{prop1} holds by Theorem \ref{thm1}.

\end{document}